\documentclass{mn2e}
\usepackage{epsfig}
\usepackage{times}
\usepackage{xfrac}
\usepackage{fix-cm}
\begin{document}

\title[Jets in hard state black holes] 
{Spectral differences between the jets in `radio loud' and `radio quiet' hard state black hole binaries}
\author[Espinasse \& Fender]
       {M. Espinasse$^{1,2}$ and R. Fender$^2$\\
       $^1$ D\'epartement de Physique, \'Ecole Normale Sup\'erieure de Cachan, 61 avenue du Pr\'esident Wilson, 94235
Cachan cedex, France \\
$^2$ Astrophysics, Department of Physics, University of Oxford, Keble Road, Oxford OX1 3RH 
\\}
\maketitle
\begin{abstract}
We have compiled from the available literature a large set of radio measurements of black hole binaries in the hard X-ray state for which measurements of the gigahertz frequency radio spectral index are possible. We separate the sample into `radio loud' and `radio quiet' subsets based upon their distribution in the radio -- X-ray plane, and investigate the distribution of radio spectral indices within each subset. 
The distribution of spectral indices of the `radio loud' subset is well described by a Gaussian distribution with mean spectral index $\alpha = +0.2$ and standard deviation $0.2$ (here spectral index is defined such that a positive spectral index means more flux at higher frequencies). The sparser sample for the `radio quiet' subset can be approximated, less well, by a Gaussian with mean $\alpha = -0.2$ and standard deviation $0.3$; alternatively the simple mean of the distribution of the radio quiet subset is $-0.3$. The two spectral index distributions are different at high statistical significance. Confirming previous work in the literature, we test to see if the differences in observed spectra could result from different distributions of jet viewing angles, but find no evidence for this. We conclude therefore that the jets in the two groups are physically different in some way, and briefly discuss possible origins and further possible diagnostics. Finally we note that extrapolating to lower frequencies the two subsets move closer together in the radio -- X-ray plane, and approximately merge into a single distribution at around 400 MHz.
\end{abstract}

\section{Introduction}

Black hole X-ray binaries (BHXRBs) exhibit a number of accretion `states', classification being based upon their X-ray spectral and fast-time variability characteristics (e.g. Done, Gierli\'nski \& Kubota 2007; Belloni \& Motta 2016). The `hard' state corresponds to an X-ray spectrum which is dominated by a component which peaks around 100 keV; the power density spectrum shows strong variability (up to 30\% rms), often with significant quasi-periodic oscillations (QPOs). The `soft' state, in contrast, has a much lower degree of variability (a few \%, with few QPOs) and an energy spectrum dominated by a component which peaks around 1 keV. The most common physical interpretation of these states is that the hard state spectrum arises from thermal comptonisation of some (poorly established) seed photon source by a hot, optically thin and geometrically thick, plasma. The soft state is modelled as a physically thin, optically thick, accretion disc whose spectrum is the integral of black body emission from each radius of the disc. Alternative explanations exist, especially those in which a jet may account for most/all of the broadband emission in the hard state (e.g. Markoff et al. 2003). Other, intermediate, accretion states exist but are rarer and usually associated with transitions between hard and soft states (or vice versa). 

Radio emission from X-ray binaries, and indeed all accreting sources, is seen as the signature of jets, which carry matter, energy and angular momentum away from the central accretor. The origin of the radio emission is synchrotron radiation from accelerated electrons in magnetic fields. Fender (2001) established that black hole X-ray binaries (BHXRBs) always show radio emission in the hard state. This radio emission was found to have a rather flat spectrum and low degree of variability compared to the more strongly peaked and variable emission observed during state transitions. In marked contrast, very little core radio emission is observed in the soft state leading to an interpretation of the absence of a jet in such states (but see e.g. Drappeau et al. 2017).
These properties were later incorporated into a more general model linking accretion states and jets (Fender, Belloni \& Gallo 2004).

A fundamental tool in studying the connection between accretion and jet production in BHXRBs has been to make quasi-simultaneous measurements in the radio and X-ray bands, and plot them together in the radio -- X-ray plane. This is particularly valuable for sources in the hard state, which usually evolves relatively slowly (days) and follows clear correlations. A strong correlation in the plane was initially found for the source GX 339--4 (Hannikainen et al. 1998) which has shown a remarkably persistent relation between radio and X-ray flux over a 15-year period which includes seven outbursts during which the jet appears to have disappeared and then reformed (Corbel et al. 2013). This correlation was initially shown to be `universal' (Gallo, Fender \& Pooley 2003) and led, with the inclusion of a mass term, to the broader works on the fundamental plane of black hole activity (Merloni, Heinz \& di Matteo 2003; Falcke, Koerding \& Markoff 2004). However, it subsequently became clear that in high-luminosity hard states some BHXRBs were considerably less luminous in the radio band than the canonical sources such as GX 339--4 (e.g. Coriat et al. 2011 and references therein). At least one of this `radio-quiet' hard state population, H1743--322, is also relatively faint in the near-infrared band (Chaty, Mu\~noz Arjonilla \& Dubus 2015).
Attempts to discern the underlying cause of the difference between the `radio loud' and `radio quiet' branches of the hard state have not been successful (e.g. Soleri \& Fender 2011; Gallo et al. 2014 and discussion therein). The only observational discriminant reported between the two branches, apart from their position in the radio -- X-ray plane, is some evidence that the radio loud sources have higher average X-ray rms variability (Din\c{c}er et al. 2014). Proposed explanations for the origin of the two tracks include differences in jet magnetic field (Casella \& Pe'er 2009), accretion flow radiative efficient (Coriat et al. 2011), or an inner accretion disc (Meyer-Hofmeister \& Meyer 2014). 

In this paper we investigate the radio spectral properties of hard state BHXRBs and find a clear difference between the radio loud and radio quiet branches. This is the first time a clear signature of physical difference between the jets in the two groups has been found.

\section{Analysis}

\begin{figure*}
\centering
\includegraphics[width=\textwidth]{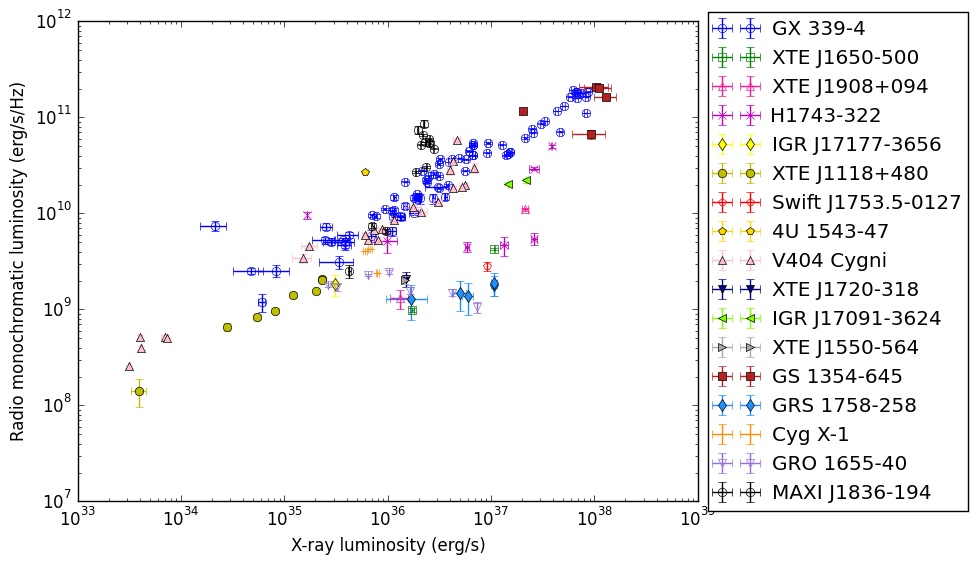}
\caption{The sample of hard state X-ray binaries used in our study, with measurements plotted in the radio -- X-ray plane. Full references for the sample are provided in the main text.}
\label{sample}
\end{figure*}

\subsection{Our sample}

Our sample consists of radio observations for 17 BHXRBs in the hard state. The radio observations used were those for which more than one frequency was available, so the spectral index could be calculated. X-ray observations at almost simultaneous times were also used when they could be found (that is the case for around 90\% of the radio observations in our data sample). By almost simultaneous we mean observations taken less than a day apart (except for the observation of XTE J1720--318, but at that period the source was not seen to be highly variable). The years for the observations used can be found in Table \ref{summary}.

\begin{figure}
\centering
\includegraphics[width=\columnwidth]{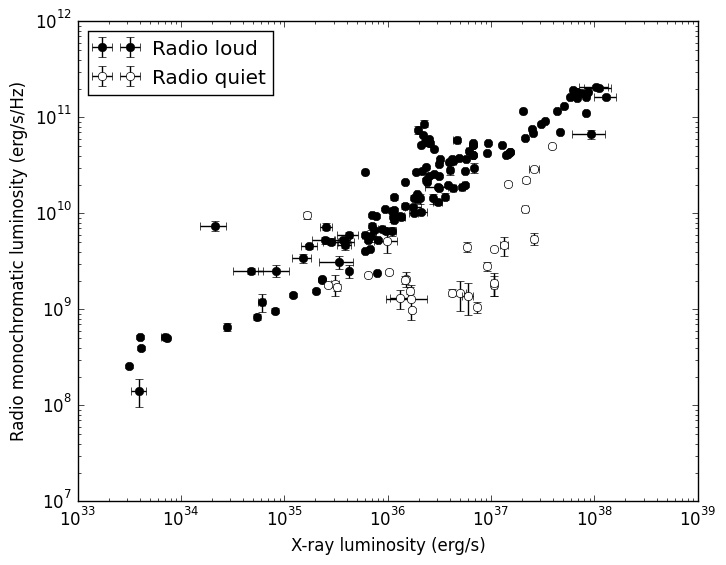}
\caption{The separation of the sample from Fig. \ref{sample} into radio loud and radio quiet branches.}
\label{classif}
\end{figure}

The BHXRBs were classified into two categories, radio loud and radio quiet, according to their position in the radio -- X-ray plane (see Figures \ref{sample} and \ref{classif}). The radio loud sources are those with a majority of points on the radio loud branch, and the radio quiet sources the ones with a majority of points on the radio quiet branch. 

The radio loud sources are: GX 339--4 (data from Corbel et al. 2000, 2013), XTE J1118+480 (data from Brocksopp et al. 2010), 4U 1543--47 (data from Kalemci et al. 2005), V404 Cygni (data from Plotkin et al. 2017, Han \& Hjellming 1992, Corbel et al. 2008), GS 1354--645 (data from Brocksopp et al. 2001, Coriat et al. 2015, Stiele \& Kong 2016) and MAXI J1836--194 (data from Jana et al. 2016, Russell et al. 2014, 2015).

The radio quiet sources are: XTE J1650--500 (data from Corbel et al. 2004, Miller et al. 2004), XTE J1908+094 (data from Curran et al. 2015), H1743--322 (data from McClintock et al. 2009, Corbel et al. 2005, Jonker et al. 2010, Coriat et al. 2011, Kalemci et al. 2006, 2008), IGR J17177--3656 (data from Paizis et al. 2011), Swift J1753.5--0127 (data from Cadolle-Bel et al. 2007), XTE J1720--318 (data from Brocksopp et al. 2005, Cadolle-Bel et al. 2004), IGR J17091--3624 (data from Rodriguez et al. 2011), XTE J1550--564 (data from Corbel et al. 2001, Tomsick et al. 2001), GRS 1758--258 (data from Lin et al. 2000),  GRO J1655--40 (data from Shaposhnikov et al. 2007, Kalemci et al. 2016). 

Cygnus X--1 (data from Pandey et al. 2006, Cadolle-Bel et al. 2006) was classified as radio loud because for the three observations we had, its spectral index was positive. However the outcome of this analysis is not affected significantly whichever group we place it into. 

\subsection{Analysis}

Once the sources are classified `loud' or `quiet', we can use all the available radio data even if there is no X-ray measurement, thus in the analysis below the plot of the spectral index against the radio luminosity has more points than the same plot against the X-ray luminosity. \\

The spectral index was either found in the papers giving the radio data, when it had been calculated by their authors, or was calculated by us when we had measurements at more than one radio frequency with the formula  

\[
\alpha = \frac{\log \left( \sfrac{F_{2}}{F_{1}}\right)}{\log \left( \sfrac{\nu _{2}}{\nu _{1}}\right)}
\]

($F$ being the radio flux density and $\nu$ the frequency). The uncertainities were calculated by propagating the errors on the fluxes, assuming there was no error on the frequencies, with the formula 

\[
err(\alpha) = \frac{1}{\ln(\sfrac{\nu_{1}}{\nu_{2}})} \sqrt{\left(\frac{err(F_{1})}{F_{1}}\right)^{2}+\left(\frac{err(F_{2})}{F_{2}}\right)^{2}} 
\]

For 87\% of the data, the spectral index was calculated between 5 GHz and 9 GHz. There apparently is no bias due to the frequencies used to calculate the spectral indices, as the remaining 13\% are not outliers in any sense.

The monochromatic radio luminosity $L_{\nu} = F_{\nu} 4 \pi d^2$, in units of erg s$^{-1}$ Hz$^{-1}$, was calculated using the best estimates for distances of the BHXRBs we could find in the literature; the numbers used are summarised in Table \ref{summary}. Russell et al. (2014) estimate the distance of MAXI J1836--194 to be between 4 kpc and 10 kpc. We decided to use the lower limit of 4 kpc as at larger distances the luminosity of MAXI J1836--194 appears quite discrepant from the other black holes, but regardless of distance its spectral indices are consistent with the rest of the loud sources.

\begin{table*}
 \centering
 \begin{tabular}{|c|c|c|c|}
  \hline
  BHXRB & Classification & Distance (kpc) & Data taken in \\ \hline
  GX 339--4 & loud & 8.4 [1] & 1996, 1999, 2002-2012 \\
  XTE J1650--500 & quiet & 2.6 [2] & 2001 \\
  XTE J1908+094 & quiet & 11 [3] & 2013-2014 \\
  H1743--322 & quiet & 8.5 [4] & 2003, 2008 \\
  IGR J17177--3656 & quiet & 8 [5] & 2011 \\ 
  XTE J1118+480 & loud & 1.72 [6] & 2005 \\
  Swift J1753.5--0127 & quiet & 6 [7] & 2005 \\
  4U 1543--47 & loud & 7.5 [8] & 2002 \\
  V404 Cygni & loud & 2.39 [9] & 1989, 2015 \\
  XTE J1720--318 & quiet & 6.5 [10] & 2003 \\ 
  IGR J17091--3624 & quiet & 11 [11] & 2011 \\
  XTE J1550--564 & quiet & 4.38 [12] & 2000 \\
  GS 1354--645 & loud & 25 [13] & 1997, 2015 \\
  GRS 1758--258 & quiet & 8.5 [14] & 1997 \\
  Cyg X--1 & loud$^{\dagger}$ & 1.86 [15] & 2003-2004 \\
  GRO J1655--40 & quiet & 3.2 [16] & 2005 \\
  MAXI J1836--194 & loud & 4$^{*}$ [17] & 2011-2012 \\ \hline
 \end{tabular}
 \caption{Our data sample. References for the distance: [1] Parker et al. 2016 [2] Homan et al. 2006 [3] Curran et al. 2015 [4] Steiner et al. 2012 [5] Paizis et al. 2011 [6] Gelino et al. 2006 [7] Cadolle-Bel et al. 2007 [8] Park et al. 2004 [9] Miller-Jones et al. 2009 [10] Chaty \& Bessolaz 2006 [11] Rodriguez et al. 2011 [12] Orosz et al. 2011a [13] Casares et al. 2009 [14] Main et al. 1999 [15] Reid et al. 2011 [16] Hjellming \& Rupen 1995 [17] Russell et al. 2014. $^{\dagger}$See text for justification. $^{*}$See text for justification. }
 \label{summary}
\end{table*}

\begin{figure}
\centering
\includegraphics[width=\columnwidth]{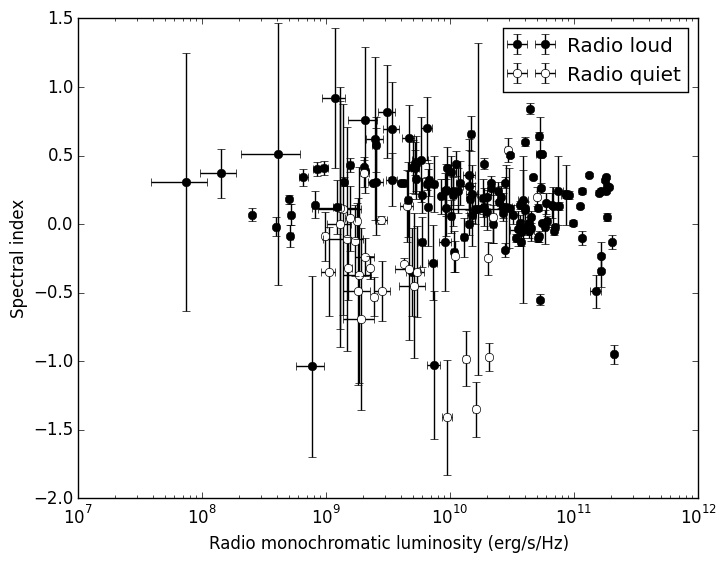}
\includegraphics[width=\columnwidth]{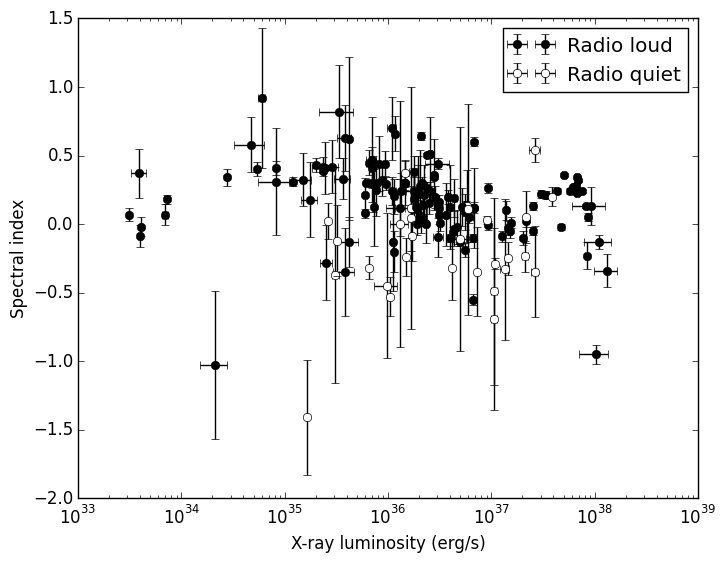}
\caption{The spectral index against monochromatic radio luminosity (top) and against 1-10 keV X-ray luminosity (bottom). }
\label{Lr alpha}
\end{figure}

Figure \ref{Lr alpha} (upper panel) displays the spectral index against the monochromatic radio luminosity. It can be seen that radio loud sources tend to have a positive spectral index, whereas radio quiet sources tend to have a negative spectral index. We studied the same correlation for the plot of spectral index against X-ray luminosity $L_X$. The X-ray luminosity was obtained using X-ray fluxes (erg cm$^{-2}$ s$^{-1}$) or photon count rates (counts s$^{-1}$) found in the literature (references above). Those fluxes or count rates were converted to unabsorbed X-ray fluxes in the 1-10 keV band using the {\sc HEASARC} tool {\sc WebPIMMS}, using the parameters $N_{H}$ and $\Gamma$ found in the same articles. When no $\Gamma$ was given, it was assumed to be 1.6 (Done et al. 2007 and references therein) as the sources are in the hard state. Then, the fluxes were converted into luminosity in erg s$^{-1}$ using the same distance estimates as for the radio monochromatic luminosity. The corresponding plot is displayed in Figure \ref{Lr alpha} (lower panel).

Although they both demonstrate the same fundamental result, namely that radio quiet sources have more `optically thin' radio spectra, it is important to show this result as a function of both (monochromatic) radio and X-ray luminosity. The reason for this is that different approaches to the underlying accretion physics may view either as the more fundamental parameter. Most approaches to accretion would treat $L_X$ as the fundamental measure, but the accretion efficiency is uncertain in the hard state, and other approaches consider the radio luminosity as a better tracer of accretion rate (e.g. Koerding, Fender \& Migliari 2006). \\

\begin{figure}
\centering
\includegraphics[width=\columnwidth]{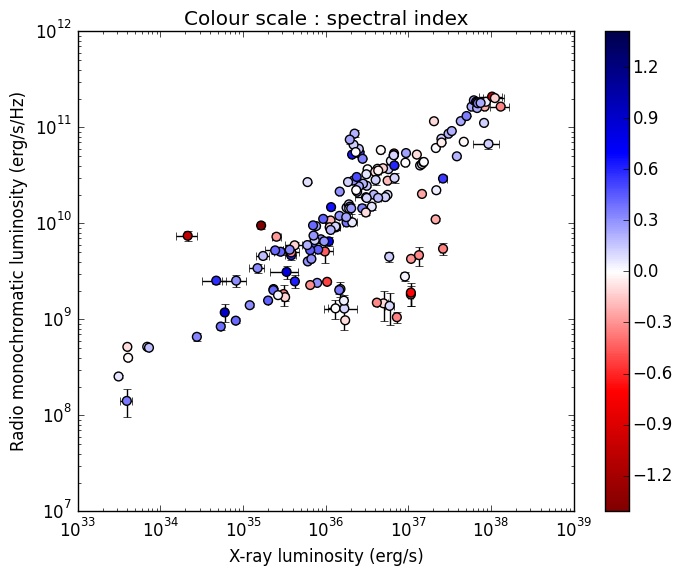}
\caption{Colour map of the spectral index in the radio -- X-ray plane.}
\label{colour}
\end{figure}

Figure \ref{colour} displays the spectral index for our data points in the radio -- X-ray plane. It demonstrates that the majority of observations have a positive spectral index in the radio loud branch, and a negative one in the radio quiet branch, for the (important) subset of measurements where there are contemporaneous radio and X-ray measurements. There are some anomalies: the two points with a spectral index of less than -1 above the radio loud branch belong to GX 339--4 (MJD 55678) and H1743--322 (MJD 52955) respectively. 
On top of the radio loud branch, the source with a negative spectral index while radio loud is GS 1354--645.
The limitation of this figure is that the error bars of the spectral index are not taken into account, and they sometimes are quite large compared to the value of the spectral index. 

\begin{figure}
\centering
\includegraphics[width=\columnwidth]{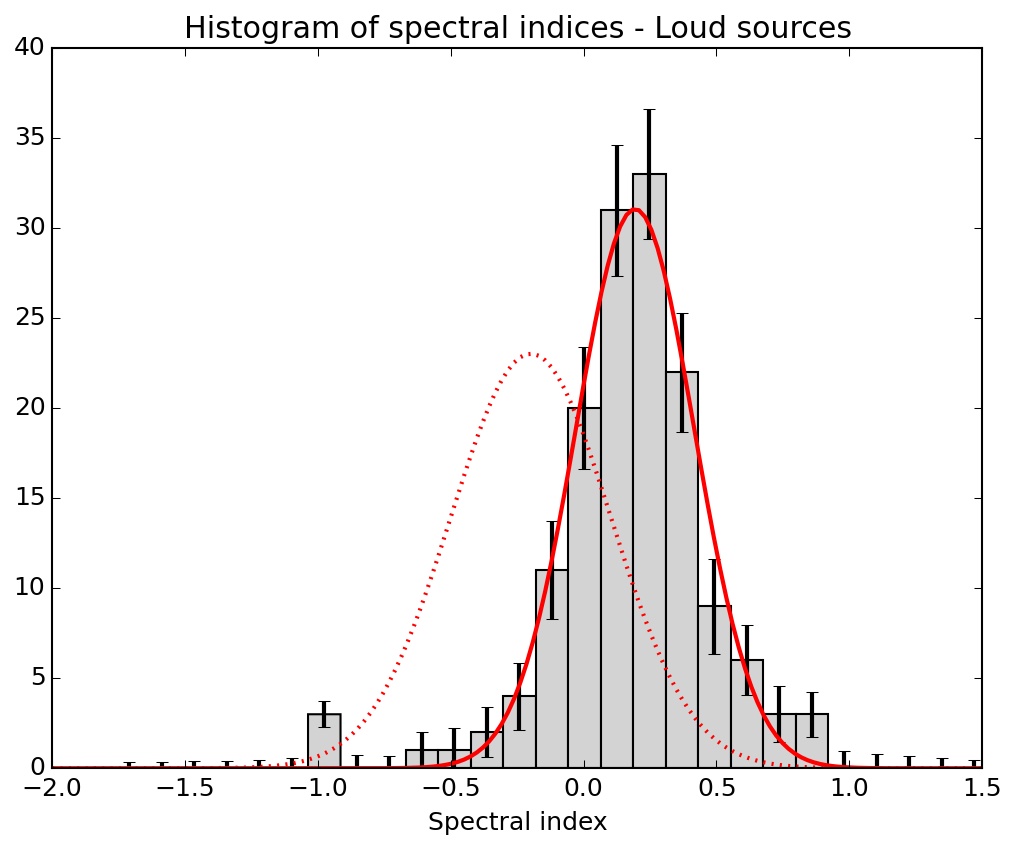}
\includegraphics[width=\columnwidth]{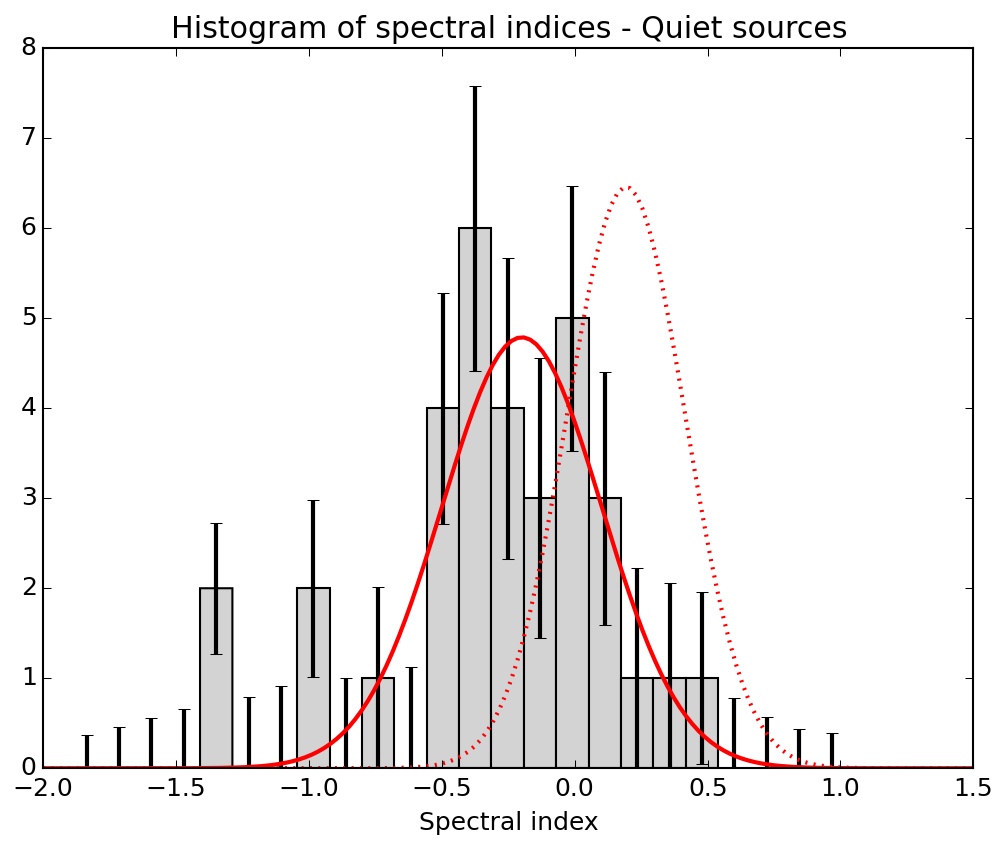}
\caption{Histogram of the spectral index for the two groups of our classification with error bars on the number of counts per bin. The red full line is a Gaussian fit of each histogram, and the dotted line is the Gaussian fit of the other group of sources, normalized to have the same integrated area as the fitted Gaussian. The radio loud sample is well fit by Gaussian centred at $\alpha = +0.2$ and it is clear that the distribution of spectral indices for the radio quiet sample is not consistent with this.}
\label{histo}
\end{figure}

In Figure \ref{histo} we present histograms of the spectral index distributions for the radio loud (upper panel) and radio quiet (lower panel) subsets. Our sample contains considerably more data for the radio loud sources than for the radio quiet ones. Nevertheless, it is clearly apparent that the radio loud sources have more positive spectral indices than the radio quiet sample.
Error bars on the number of counts for each bin of the histograms were calculated as follows: we assume that the errors on the spectral index measurements are normally distributed, with the error bars representing $1 \sigma$; the variance in each bin was then calculated using Bernoulli variables.

We fitted the histograms with Gaussian functions to test the difference in the spectral index distributions between the radio-loud and radio-quiet subsets. The fit function was

\[
gauss(x) = a \exp{ \left( -\frac{(x-b)^{2}}{2c^{2}} \right) }
\]

where $b$ is the expectation value and $c$ is the standard deviation for the distribution.

Fitting took into account the error bars of the histograms and gave the following coefficients: for loud sources, $a = 31.1 \pm 2.3$, $b = 0.19 \pm 0.02$ and $c = 0.22 \pm 0.01$, and for quiet sources, $a = 4.79 \pm 0.89$, $b = -0.20 \pm 0.05$ and $c = 0.30 \pm 0.05$. The fit of a Gaussian function is good for the radio loud sources; for the radio quiet population it is less so, but still adequate. Based upon these fits, the radio-loud sources have a more positive expectation value and a slightly narrower distribution. To test if the two fitted Gaussian distributions were statistically different, we performed a Welch two sample t-test using the programming language {\sc R}. The p-value of $1.8 \times 10^{-6}$ (confidence interval of 95\%) confirms that the mean values for the two Gaussians are significantly different. 

\begin{figure}
\centering
\includegraphics[width=\columnwidth]{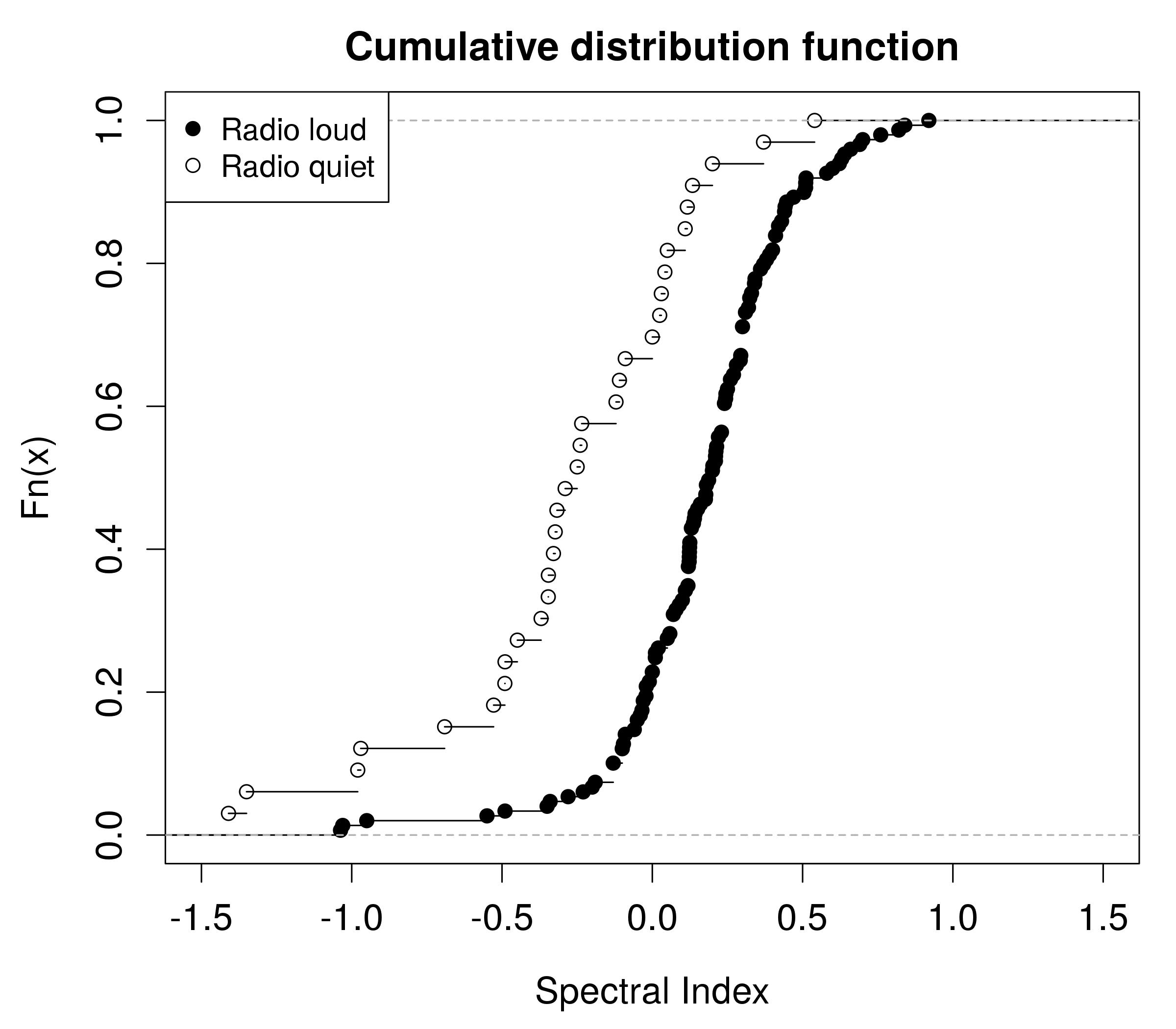}
\caption{Cumulative distribution function for the spectral indices of the radio loud and radio quiet samples.} 
\label{CDF}
\end{figure}

In an alternative approach, we have calculated the cumulative distribution functions of the two samples (Figure \ref{CDF}). The samples again appear to be convincingly different. Finally we performed a Kolmogorov--Smirnov test; the p-value of $2.4 \times 10^{-7}$ obtained confirms that the two samples come from different distributions.

\begin{figure}
\centering
\includegraphics[width=\columnwidth]{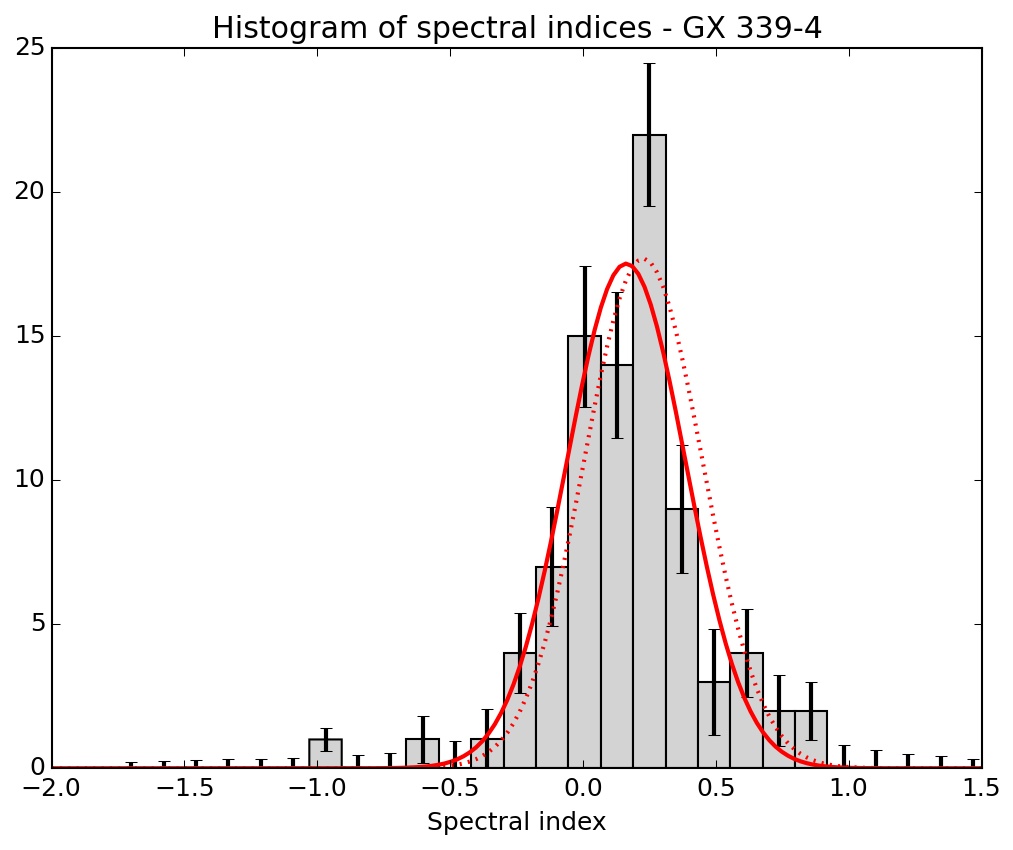}
\includegraphics[width=\columnwidth]{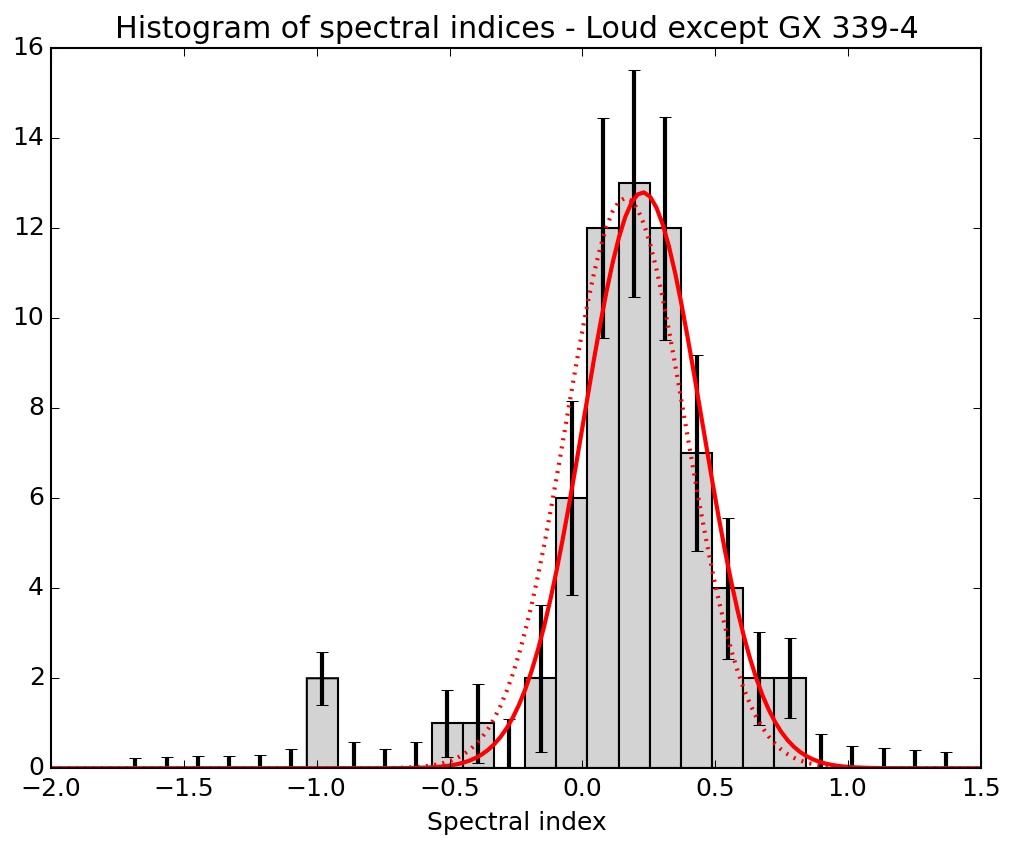}
\caption{Same as Fig. \ref{histo} but with GX 339--4 on the upper panel and all radio loud sources except GX 339--4 on the lower panel. Both samples are well fit by Gaussian centred at $\alpha = +0.2$.}
\label{339}
\end{figure}

Since GX 339--4 has more data than any other source in our sample, we investigated whether it may be biasing the results for the radio loud sample.  In Fig. \ref{339} we have plotted two histograms of the spectral index: one using only GX 339--4 data points (Fig. \ref{339} upper panel), and the other one using all radio loud sources except GX 339--4 (Fig. \ref{339} lower panel). The error bars on the counts per bin were calculated as explained before. We then fitted both histograms with Gaussians. The parameters for the GX 339--4 Gaussian are $a = 17.5 \pm 1.6$, $b = 0.16 \pm 0.02$ and $c = 0.22 \pm 0.02$, and for the other loud sources they are $a = 12.8 \pm 1.5$, $b = 0.23 \pm 0.02$, $c = 0.22 \pm 0.02$. These values are very similar, and are also very close to the parameters found for the Gaussian fitting the whole loud distribution. It can be seen that both distributions seem well fitted by Gaussians, thus the Gaussian shape was not due only to GX 339--4 dominating the sample. To confirm that the two Gaussians are statistically similar, we performed a Welch t-test as previously, and we obtained a p-value of $0.6019$. This means the difference in mean value of the Gaussians is statistically non significant. Thus, it confirms that the Gaussian shape of the radio loud sample was due to all sources, and not only GX 339--4, and furthermore that the distribution of spectral indices for GX 339--4 appears to be typical for radio loud hard state sources.

\section{Discussion}

We have shown that in the 5--9 GHz frequency range the distributions of radio spectra for the two samples are markedly different, in the sense that the radio quiet sample are more negative. 
We consider two explanations for this result: either there is a significant inclination angle dependence, and the radio loud and radio quiet sources have different distributions of inclinations, or the jets in the two samples are physically different in some way. 

\begin{figure}
\centering
\includegraphics[width=\columnwidth]{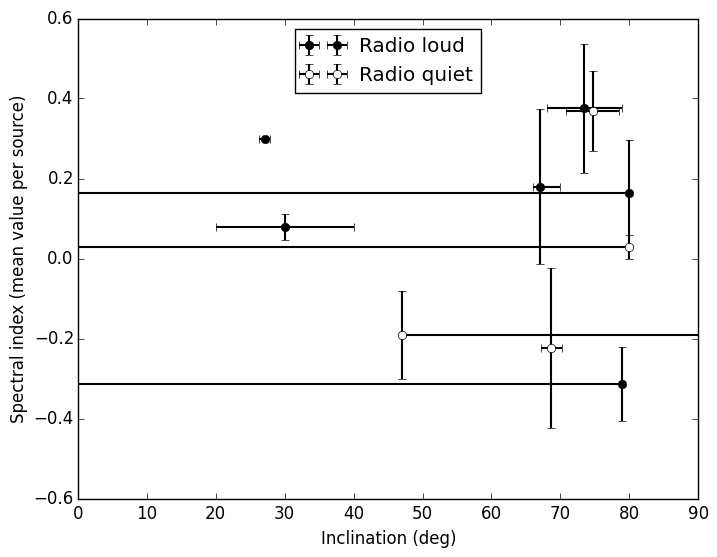}
\caption{Spectral index as a function of inclination for BHXRBs. The error bars show the intervals of possible inclination and the markers are used to indicate the nature of the source. The radio loud source with $\bar{\alpha} \simeq -0.3$ is GS 1354--645 (5 observations) and the radio quiet source with $\bar{\alpha} \simeq 0.37$ is XTE J1550--564 (1 observation). There is no apparent connection between radio loudness and inclination. }
\label{inclin}
\end{figure}

Addressing first the question of inclination, some models for relativistic jets have a viewing angle dependence for the spectral index (e.g. Falcke 1996; see Fig. 2 of Markoff et al. 2003). Although this has been previously investigated (e.g. Soleri \& Fender 2011) we consider again here the possibility that inclinations could be responsible for the separation of the groups. The inclinations of most BHXRBs are not very well constrained, and often only upper or lower limits are available. Figure \ref{inclin} illustrates the different inclinations for the two groups of sources (see Table \ref{incl} for the inclinations and their references). We see that both radio loud and radio quiet sources can have a small or large inclination. Thus, it is not likely that the difference in slope of the spectra is explained by the viewing angle of the jets. However, if the inclination of the jet really is the origin of the difference in spectral indices, it implies that the jets are strongly misaligned with the orbits (see e.g. discussion in Martin, Tout \& Pringle 2008). \\

If inclination is not the cause, what is the underlying physical reason for the spectral index difference? The obvious first conclusion to draw is that the radio loud sources are more optically thick than the radio quiet sources. The mean spectral index for the radio loud sources is consistent with that expected for models of partially self-absorbed jets, probably reheated via internal shocks, where the radio emission all arises at frequencies below the break from optically thick to optically thin  (Blandford \& K\"{o}nigl 1979, Malzac 2013). The radio quiet sample includes spectral indices which are consistent with completely optically thin emission ($\alpha = -0.55$ for a lepton spectrum index of $p=2.1$). However, the mean value of $\simeq -0.3$ suggests a mix of optically thin and thick emission and/or ejecta which are evolving between thick and thin states (i.e. discrete ejecta). In the following subsections we investigate first whether there is any evidence for a dependence on viewing angle, and then briefly discuss other possible explanations.

\subsection{Dependence on accretion flow?}

If the emission from the radio quiet sources is indeed composed of more discrete ejecta, we might expect the radio (and presumably also X-ray) emission to be more strongly variable and separable into individual flares (as observed in e.g. GRS 1915+105). Current data sets are rather sparse, but there is no evidence in the radio data for more variability in either group of sources. The result of Din\c{c}er et al. (2014), who report that there is more X-ray variability in the radio loud objects (`standards' in their terminology) deserves more attention and should be followed up.

\subsection{Low-frequency break to optically thin emission?}

Alternatively, the radio quiet sources could be those for which the break from optically thick (flat spectrum) to thin emission in the compact jet is in the radio band. However, for most studied BHXRBs this break has been estimated to occur in the infrared band in the hard state (e.g. Corbel \& Fender 2002; Migliari et al. 2010; Russell et al. 2013). If this were the correct interpretation, it would require the break frequency to be a factor $\sim 10^5$ different between the two branches, which seems unlikely. 

\subsection{Jet-ISM interaction}

It is well observed that radio emission arising from the interaction of BHXRB jets with the ambient medium are more optically thin than the core jets (e.g. Corbel et al. 2002). However, in order for a the sources to be {\em less} radio luminous whilst also generating an extra component of radio emission would require the core jets themselves to be less powerful, which brings us back to different core physics.

\subsection{Spin?}

\begin{figure}
\centering
\includegraphics[width=\columnwidth]{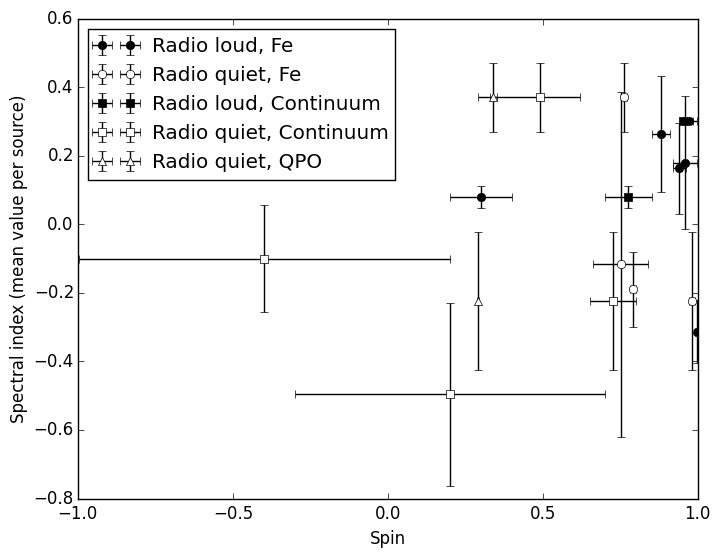}
\caption{Spectral index as a function of estimated black hole spin. The three methods of spin estimation are represented with different symbols while the colour represents the nature of the source. There is no apparent connection between radio loudness and spin estimate.}
\label{spins}
\end{figure}

We have also studied the possibility that the difference in spectral index between the radio loud and radio quiet sources was due to a difference in the black hole spin for those sources. We gathered spin estimates calculated using three methods: the fit of the Fe line ('reflection'), the fit of the continuum of the disc spectrum, and the QPOs of the sources. These spin estimates and their sources can be found in Table \ref{spin}.
The spin values were plotted against the mean spectral index for each source on Figure \ref{spins}. There is no obvious connection between the reported spin measurements and which radio track an obect is on, although the samples are very small. This implies that either the spin does not explain the differences we observe in spectral indices, or the spin measurements themselves are incorrect.

\subsection{Extrapolation to lower frequencies}

\begin{figure}
\centering
\includegraphics[width=\columnwidth]{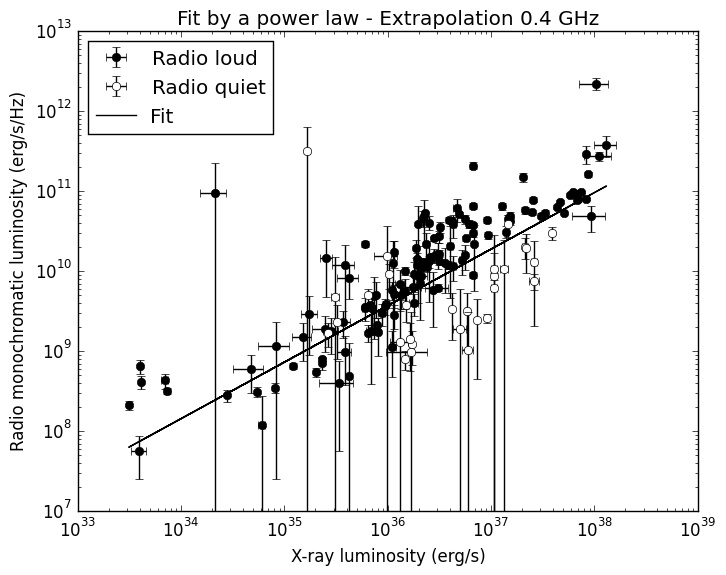}
\caption{Extrapolation of radio:X-ray plane to 400 MHz, using the measured spectral indices, and a fit to a single power law ($L_{R} \propto L_{X}^{0.71}$). At this frequency the presence of two tracks would not be apparent. }
\label{extrapol}
\end{figure}

Regardless of the underlying physics, the signs of the different spectra in the two groups suggest that extrapolating to lower frequencies should bring the two populations closer together. We have performed this extrapolation (Figure \ref{extrapol}) to a frequency of 400 MHz. We were limited by the large error bars on the spectral index: when we extrapolate to lower frequencies, the error bars on the luminosity become larger than the scatter between the groups. At 400 MHz a reasonable fit can be achieved to a {\em single} power law through the combined sample. Clearly in Figure \ref{extrapol} the radio quiet sources are still below the loud branch, but within uncertainties the branches have effectively merged. The fit is as follows: $\log(L_{R}) = a L_{X} + b$ with $a = 0.71 \pm 0.02$ and $b = -15.86 \pm 0.86$; this is a slope very similar to that overall for the radio-loud hard state sources alone (perhaps not surprising as they dominate the merged population). 

\begin{table}
 \centering
 \begin{tabular}{|l|l|l|}
  \hline
  BHXRB & Inclination (deg) & Reference \\ \hline
  GX 339--4 & i $\leq$ 80 & No X-ray eclipses$^\dagger$ \\ 
  XTE J1650--500 & i $\geq$ 47 & [1] \\
  XTE J1908+094 & \multicolumn{2}{l}{No dynamical information available} \\
  H1743--322 & \multicolumn{2}{l}{No dynamical information available} \\
  IGR J17177--3656 & \multicolumn{2}{l}{No dynamical information available} \\
  XTE J1118+480 & 68 $\leq$ i $\leq$ 79 & [2] \\
  Swift J1753.5--0127 & i $\leq$ 80 & No X-ray eclipses$^\dagger$ \\ 
  4U 1543--47 & 20 $\leq$ i $\leq$ 40 & [3] \\ 
  V404 Cygni & 67$^{+3}_{-1}$ & [4] \\ 
  XTE J1720--318 & \multicolumn{2}{l}{No dynamical information available} \\
  IGR J17091--3624 & \multicolumn{2}{l}{No dynamical information available} \\
  XTE J1550--564 & 74.7 $\pm$ 3.8 & [5] \\
  GS 1354--645 & i $\leq$ 79 & [6] \\
  GRS 1758--258 & \multicolumn{2}{l}{No dynamical information available} \\
  Cyg X--1 & 27.1 $\pm$ 0.8 & [7] \\
  GRO J1655--40 & 68.7 $\pm$ 1.5 & [8] \\
  MAXI J1836--194 & \multicolumn{2}{l}{No dynamical information available} \\
  \hline
 \end{tabular}
 \caption{The inclinations derived from fits to ellipsoidal light curves found in the literature. $^\dagger$\emph{Source: J. Casares}. References: [1] Orosz et al. 2004 [2] Khargharia et al. 2013 [3] Orosz et al. 1998 [4] Khargharia et al. 2010 [5] Orosz et al. 2011a [6] Casares et al. 2009 [7] Orosz et al. 2011b [8] Beer \& Podsiadlowsky 2002.}
 \label{incl}
\end{table}

\begin{table}
 \centering
 \begin{tabular}{|l|l|l|c|}
  \hline
  BHXRB & Spin & Method & Reference \\ \hline
  GX 339--4 & 0.94 $\pm$ 0.02 & Fe line & [1] \\ 
  XTE J1650--500 & 0.79 $\pm$ 0.01 & Fe line  & [1]  \\
  XTE J1908+094 & 0.75 $\pm$ 0.09 & Fe line  & [1] \\
  H1743--322 & 0.2 $\pm$ 0.5 & Disc continuum & [2] \\
  IGR J17177--3656 & \multicolumn{3}{l}{No spin estimate found in the literature}\\
  XTE J1118+480 & \multicolumn{3}{l}{No spin estimate found in the literature} \\
  Swift J1753.5--0127 & \multicolumn{3}{l}{No spin estimate found in the literature} \\ 
  4U 1543--47 & 0.3 $\pm$ 0.1 & Fe line & [1] \\
   & 0.7 $\leq$ a $\leq$ 0.85 & Disc continuum & [3] \\
  V404 Cygni & a $\geq$ 0.92 & Fe line & [4] \\ 
  XTE J1720--318 & \multicolumn{3}{l}{No spin estimate found in the literature} \\
  IGR J17091--3624 & a $\leq$ 0.2 & Disc continuum & [5] \\
  XTE J1550--564 & 0.76 $\pm$ 0.01 & Fe line & [1] \\
   & 0.49$^{+0.13}_{-0.2}$ & Fe + Disc cont. & [6] \\
   & 0.34 $\pm$ 0.01 & QPO & [7] \\
  GS 1354--64 & 0.998$_{-0.009}$ & Fe line & [8] \\
  GRS 1758--258 & \multicolumn{3}{l}{No spin estimate found in the literature} \\
  Cyg X--1 & 0.97$^{+0.014}_{-0.02}$ & Fe line & [9] \\
   & a $\geq$ 0.95 & Disc continuum & [10] \\
  GRO J1655--40 & 0.98 $\pm$ 0.01 & Fe line & [1] \\
   & 0.65 $\leq$ a $\leq$ 0.8 & Disc continuum & [3] \\
   & 0.29 $\pm$ 0.003 & QPO & [11] \\
  MAXI J1836--194 & 0.88 $\pm$ 0.03 & Fe line & [12] \\
  \hline
 \end{tabular}
 \caption{The spins of the BHXRBs and the method used to get them. References: [1] Miller et al. 2009 [2] Steiner et al. 2012 [3] Shafee et al. 2006 [4] Walton et al. 2017 [5] Rao \& Vadawale 2012 [6] Steiner et al. 2011 [7] Motta et al. 2014b [8] El-Batal et al. 2016 [9] Fabian et al. 2012 [10] Gou et al. 2011 [11] Motta et al. 2014a [12] Reis et al. 2012.}
 \label{spin}
\end{table}

\section{Conclusions}

We have demonstrated that radio loud black hole X-ray binaries have significantly different radio spectra than their radio quiet counterparts, in a spectral regime expected to be dominated by a core jet. Ruling out an origin in inclination effects (unless the jets are highly misaligned with the orbital planes), we conclude that the physics of the core jets is different between the two samples. We encourage further observations, such as detailed comparitive studies of X-ray, radio and infrared variability on the two branches, and possibly also polarisation, in order to better understand this phenomenon.

\section*{Acknowledgements}

We would like to thank Dave Russell, Sara Motta, Jorge Casares, Sera Markoff and in particular St\'ephane Corbel for valuable comments on a draft of this paper. 
RF was partly funded by ERC Advanced Investigator Grant 267607 ‘4 PI SKY’.


\begin{thebibliography}{}

\bibitem[]{beer02}
Beer M.E., Podsiadlowski P., 2002, MNRAS, 331, 351

\bibitem[]{belloni16}
Belloni T.M., Motta S.E., 2016, Astrophysics of Black Holes, Astrophysics and Space Science Library, Volume 440


\bibitem[]{blandford79}
Blandford R., K\"{o}nigl A., 1979, ApJ, 232, 34

\bibitem[]{brocksopp01} 
Brocksopp C., Jonker P.G., Fender R.P., Groot P.J., van der Klis M., Tingay S.J., 2001, MNRAS, 323, 517

\bibitem[]{brocksopp05} 
Brocksopp C., Corbel S., Fender R.P., Rupen M., Sault R., Tingay S.J., Hannikainen D., O'Brien K., 2005, MNRAS, 356, 125

\bibitem[]{brocksopp10} 
Brocksopp C., Jonker P.G., Maitra D., Krimm H.A., Pooley G.G., Ramsay G., Zurita C., 2010, MNRAS, 404, 908

\bibitem[]{cadollebel04} 
Cadolle Bel M. et al., 2004, A\&A, 426, 659 

\bibitem[]{cadollebel06} 
Cadolle Bel M. et al., 2006, A\&A, 446, 591 

\bibitem[]{cadollebel07} 
Cadolle Bel M. et al., 2007, ApJ, 659, 549 

\bibitem[]{casares09} 
Casares J. et al., 2009, ApJS, 181, 238 

\bibitem[]{casella09}
Casella P., Pe'er A., 2009, ApJ, 703, L63

\bibitem[]{chaty06}
Chaty S., Bessolaz N., 2006, A\&A, 455, 639

\bibitem[]{chaty15}
Chaty S., Mu\~noz Arjonilla A.J., Dubus G., 2015, A\&A

\bibitem[]{corbel00} 
Corbel S., Fender R.P., Tzioumis A.K., Nowak M., McIntyre V., Durouchoux P., Sood R., 2000, A\&A, 359, 251

\bibitem[]{corbel01} 
Corbel S., Kaaret P., Jain R.K., Bailyn C.D., Fender R.P., Tomsick J.A., Kalemci E., McIntyre V., Campbell-Wilson D., Miller J.M., McCollough M.L., 2001, ApJ, 554, 43

\bibitem[]{corbel02a}
Corbel S., Fender R., 2002, ApJ, 573, L35

\bibitem[]{corbel02b}
Corbel S., Fender R.P., Tzioumis A.K., Tomsick J.A., Orosz J.A., Miller J.M., Wijnands R., Kaaret P., 2002, Science, 298, 196

\bibitem[]{corbel04}
Corbel S., Fender R.P., Tomsick J.A., Tzioumis A.K., Tingay S., 2004, ApJ, 617, 1272

\bibitem[]{corbel05} 
Corbel S., Kaaret P., Fender R.P., Tzioumis A.K., Tomsick J.A., Orosz J.A., 2005, ApJ, 632, 504

\bibitem[]{corbel08} 
Corbel S., Koerding E., Kaaret P., 2008, MNRAS, 389, 1697

\bibitem[]{corbel13} 
Corbel S., Coriat M., Brocksopp C., Tzioumis A.K., Fender R.P., Tomsick J.A., Buxton M.M., Bailyn C.D., 2013, MNRAS, 428, 2500

\bibitem[]{coriat11} 
Coriat M., Corbel S., Prat L., Miller-Jones J.C.A., Cseh D., Tzioumis A.K., Brocksopp C., Rodriguez J., Fender R.P., Sivakoff G.R., 2011, MNRAS, 414, 677

\bibitem[]{coriat15} 
Coriat M., Tzioumis T., Corbel S., Fender R., Miller-Jones J., 2015, The Astronomer's Telegram, 7656

\bibitem[]{R14}
Crawley, M.J., 2014, Statistics: an introduction using R. John Wiley \& Sons, p. 145-146

\bibitem[]{curran15} 
Curran P.A. et al., 2015, MNRAS, 451, 3975 

\bibitem[]{dincer14}
Din\c{c}er T., Kalemci E., Tomsick J.A., Buxton M.M., Bailyn C.D., 2014, ApJ, 795, 74

\bibitem[]{done07}
Done C., Gierli\'nski M., Kubota A., 2007, A\&ARv, 15, 1

\bibitem[]{drappeau17}
Drappeau S., et al., 2017, MNRAS, 466, 4272

\bibitem[]{el-batal16}
El-Batal A.M. et al., 2016, ApJ, 826, L12 

\bibitem[]{fabian12}
Fabian A.C. et al., 2012, MNRAS, 424, 217 

\bibitem[]{falcke96}
Falcke, H., 1996, ApJ, 464, L67

\bibitem[]{falcke04}
Falcke H., Koerding E., Markoff S., 2004, A\&A, 414, 895

\bibitem[]{fender01}
Fender R.P., 2001, MNRAS, 322, 31

\bibitem[]{fender04}
Fender R.P., Belloni T.M., Gallo E., 2004, MNRAS, 355, 1105

\bibitem[]{gallo03}
Gallo E., Fender R.P., Pooley G.G., 2003, MNRAS, 344, 60

\bibitem[]{gallo14}
Gallo E., Miller-Jones J.C.A., Russell D.M., Jonker P.G., Homan J., Plotkin R.M., Markoff S., Miller B.P., Corbel S., Fender R.P., MNRAS, 2014, 445, 290

\bibitem[]{gelino06}
Gelino D.M., Balman {\c S}., K{\i}z{\i}lo{\v g}lu, {\"U}., Y{\i}lmaz, A., Kalemci E., Tomsick J.A., 2006, ApJ, 642, 438

\bibitem[]{gou11}
Gou L., McClintock J.E., Reid M.J., Orosz J.A., Steiner J.F., Narayan R., Xiang J., Remillard R.A., Arnaud K.A., Davis S.W., 2011, ApJ, 742, 85

\bibitem[]{han92} 
Han X., Hjellming R.M., 1992, ApJ, 400, 304

\bibitem[]{hannikainen98}
Hannikainen D., Hunstead R.W., Campbell-Wilson D., Sood R.K., 1998, A\&A, 337, 460

\bibitem[]{hjellming95} 
Hjellming R.M., Rupen M.P., 1995, Nature, 375, 464

\bibitem[]{homan06} 
Homan J., Wijnands R., Kong A., Miller J.M., Rossi S., Belloni T., Lewin W.H.G., 2006, MNRAS, 366, 235

\bibitem[]{jana16}
Jana A., Debnath D., Chakrabarti S.K., Mondal S., Molla A.A., 2016, ApJ, 819, 107

\bibitem[]{jonker10} 
Jonker P.G. et al., 2010, MNRAS, 401, 1255 

\bibitem[]{kalemci05} 
Kalemci E., Tomsick J.A., Buxton M.M., Rothschild R.E., Pottschmidt K., Corbel S., Brocksopp C., Kaaret P., 2005, ApJ, 622, 508

\bibitem[]{kalemci06} 
Kalemci E., Tomsick J.A., Rothschild R.E., Pottschmidt K., Corbel S., Kaaret P., 2006, ApJ, 639, 340

\bibitem[]{kalemci08} 
Kalemci E., Tomsick J.A., Corbel S., Tzioumis T., 2008, The Astronomer's Telegram, 1378

\bibitem[]{kalemci16}
Kalemci E., Begelman M.C., Maccarone T.J., Din{\c c}er T., Russell T.D., Bailyn C., Tomsick J.A., 2016, MNRAS, 463, 615

\bibitem[]{khargharia10}
Khargharia J., Froning C.S., Robinson E.L., 2010, ApJ, 716, 1105

\bibitem[]{khargharia13}
Khargharia J., Froning C.S., Robinson E.L., Gelino D.M., 2013, AJ, 145, 21

\bibitem[]{koerding06}
Koerding E.G., Fender R.P., Migliari S., 2006, MNRAS, 369, 1451

\bibitem[]{lin00} 
Lin D., Smith I.A., Liang E.P., Bridgman T., Smith D.M., Mart{\' i} J., Durouchoux P., Mirabel I.F., Rodr{\' i}guez L.F., 2000, ApJ, 532, 548

\bibitem[]{main99}
Main D.S., Smith D.M., Heindl W.A., Swank J., Leventhal M., Mirabel I.F., Rodr{\'{\i}}guez L.F., 1999, ApJ, 525, 901

\bibitem[]{malzac13}
Malzac J., 2013, MNRAS, 429, L20

\bibitem[]{markoff03}
Markoff S., Nowak M., Corbel S., Fender R., Falcke H., 2003, A\&A 397, 645

\bibitem[]{martin08}
Martin R.G., Tout C.A., Pringle J.E., 2008, 387, 188

\bibitem[]{mcclintock09}
McClintock J.E., Remillard R.A., Rupen M.P., Torres M.A.P., Steeghs D., Levine A.M., Orosz J.A., 2009, ApJ, 698, 1398

\bibitem[]{merloni03}
Merloni A., Heinz S., di Matteo T., 2003, MNRAS, 345, 1057

\bibitem[]{meyer-hofmeister14}
Meyer-Hofmeister E., Meyer F., 2014, A\&A, 562, 142

\bibitem[]{migliari10}
Migliari S., et al., 2010, ApJ, 710, 117

\bibitem[]{miller04} 
Miller J.M., Raymond J., Fabian A.C., Homan J., Nowak M.A., Wijnands R., van der Klis M., Belloni T., Tomsick J.A., Smith D.M., Charles P.A., Lewin W.H.G., 2004, ApJ, 601, 450

\bibitem[]{miller09}
Miller J.M., Reynolds C.S., Fabian A.C., Miniutti G., Gallo L.C., 2009, ApJ, 697, 900

\bibitem[]{millerjones09} 
Miller-Jones J.C.A., Jonker P.G., Dhawan V., Brisken W., Rupen M.P., Nelemans G., Gallo E., 2009, ApJ, 706, L230

\bibitem[]{motta14a}
Motta S.E., Belloni T.M., Stella L., Mu{\~n}oz-Darias T., Fender R., 2014a, MNRAS, 437, 2554

\bibitem[]{motta14b}
Motta S.E., Mu{\~n}oz-Darias T., Sanna A., Fender R., Belloni T., Stella L., 2014b, MNRAS, 439, L65
    
\bibitem[]{orosz98}
Orosz J.A., Jain R.K., Bailyn C.D., McClintock J.E., Remillard R.A., 1998, ApJ, 499, 375

\bibitem[]{orosz04}
Orosz J.A., McClintock J.E., Remillard R.A., Corbel S., 2004, ApJ, 616, 376

\bibitem[]{orosz11a}
Orosz J.A., Steiner J.F., McClintock J.E., Torres M.A.P., Remillard R.A., Bailyn C.D., Miller J.M., 2011a, ApJ, 730, 75

\bibitem[]{orosz11b} 
Orosz J.A., McClintock J.E., Aufdenberg J.P., Remillard R.A., Reid M.J., Narayan R., Gou L., 2011b, ApJ, 742, 84

\bibitem[]{paizis11} 
Paizis A., Nowak M.A., Wilms J., Chaty S., Corbel S., Rodriguez J., Del Santo M., Ubertini P., Chini R., 2011, ApJ, 738, 183

\bibitem[]{pandey06} 
Pandey M., Rao A.P., Pooley G.G., Durouchoux P., Manchanda R.K., Ishwara-Chandra C.H., 2006, A\&A, 447, 525

\bibitem[]{park04} 
Park S.Q., Miller J.M., McClintock J.E., Remillard R.A., Orosz J.A., Shrader C.R., Hunstead R.W., Campbell-Wilson D., Ishwara-Chandra C.H., Rao A.P., Rupen M.P., 2004, ApJ, 610, 378

\bibitem[]{parker16} 
Parker M.L. et al., 2016, ApJ, 821, L6 

\bibitem[]{plotkin17} 
Plotkin R.M., 2017, ApJ, 834, 104 

\bibitem[]{R}
R Core Team (2014). R: A language and environment for statistical computing. R Foundation for Statistical Computing, Vienna, Austria. URL http://www.R-project.org/.

\bibitem[]{rao12}
Rao A., Vadawale S.V., 2012, ApJ, 757, L12

\bibitem[]{reid11} 
Reid M.J., McClintock J.E., Narayan R., Gou L., Remillard R.A., Orosz J.A., 2011, ApJ, 742, 83

\bibitem[]{reis12}
Reis R.C., Miller J.M., Reynolds M.T., Fabian A.C., Walton D.J., 2012, ApJ, 751, 34

\bibitem[]{rodriguez11} 
Rodriguez J., Corbel S., Caballero I., Tomsick J.A., Tzioumis T., Paizis A., Cadolle Bel M., Kuulkers E., 2011, A\&A, 533, L4

\bibitem[]{russell13}
Russell D.M., et al., 2013, MNRAS, 429, 815

\bibitem[]{russell14}
Russell T.D., Soria R., Miller-Jones J.C.A., Curran P.A., Markoff S., Russell D.M., Sivakoff G.R., 2014, MNRAS, 439, 1390

\bibitem[]{russell15}
Russell T.D., et al., 2015, MNRAS, 450, 1745
 
\bibitem[]{shafee06}
Shafee R., McClintock J.E., Narayan R., Davis S.W., Li L.-X., Remillard R.A., 2006, ApJ, 636, L113

\bibitem[]{shaposhnikov07} 
Shaposhnikov N., Swank J., Shrader C.R., Rupen M., Beckmann V., Markwardt C.B., Smith D.A., 2007, ApJ, 655, 434

\bibitem[]{soleri11}
Soleri P., Fender R., 2011, MNRAS, 413, 2269

\bibitem[]{steiner11}
Steiner J.F., Reis R.C., McClintock J.E., Narayan R., Remillard R.A., Orosz J.A., Gou L., Fabian A.C., Torres M.A.P., 2011, MNRAS, 416, 941

\bibitem[]{steiner12} 
Steiner J.F., McClintock J.E., Reid M.J., 2012, ApJ, 745, L7

\bibitem[]{stiele16} 
Stiele H., Kong A.K.H., 2016, MNRAS, 459, 4038

\bibitem[]{tomsick01} 
Tomsick J.A., Corbel S., Kaaret P., 2001, ApJ, 563, 229

\bibitem[]{walton17}
Walton D.J. et al., 2017, ApJ, 839, 110 

\end{thebibliography}
\end{document}